\documentclass[
  aps,
  prd,
  twocolumn,
  preprintnumbers,
  nofootinbib,
  superscriptaddress,
  floatfix,
  amsmath
]{revtex4-2}

\usepackage[english]{babel}
\usepackage[utf8]{inputenc}
\usepackage{amsmath,amssymb,bm}
\usepackage{graphicx}
\usepackage{microtype}
\usepackage{hyperref}
\usepackage[normalem]{ulem} 

\graphicspath{{figs/}{./}}
\DeclareGraphicsExtensions{.pdf,.png,.jpg,.jpeg}
\setkeys{Gin}{draft=false}

\hypersetup{
  colorlinks=true,
  linkcolor=blue,
  citecolor=blue,
  urlcolor=blue,
  pdftitle={Effective Neutrino Mass and Apparent Phantom Crossing from a pNGB Mass-Varying-Neutrino Model},
  pdfauthor={Gabriel Rodrigues}
}

\begin{document}

\title{Effective Neutrino Mass and Apparent Phantom Crossing}

\author{Gabriel Rodrigues}
\email{gabrielrodrigues@on.br}
\affiliation{Observatório Nacional, Rio de Janeiro - RJ, 20921-400, Brasil}
\author{Jamerson Rodrigues}
\email{ jamerson.rodrigues@ufrn.br
}
\affiliation{Departamento de Física Teórica e Experimental,
Universidade Federal do Rio Grande do Norte, Natal - RN, 59072-970, Brasil}
\affiliation{Observatório Nacional, Rio de Janeiro - RJ, 20921-400, Brasil}
\author{Simony Santos da Costa}
\email{simony.costa@fisica.ufrn.br}
\affiliation{Instituto Internacional de Física, Universidade Federal do Rio Grande do Norte, Natal - RN, 59072-970, Brasil}
\author{Jailson Alcaniz}
\email{alcaniz@on.br}
\affiliation{Observatório Nacional, Rio de Janeiro - RJ, 20921-400, Brasil}

\date{\today}

\begin{abstract}

Recent cosmological constraints on the sum of neutrino masses have reached, and in some analyses fallen below, the minimum value implied by the normal mass ordering, suggesting a possible tension with neutrino-oscillation measurements. We investigate whether this discrepancy can be alleviated in a mass-varying-neutrino (MaVaN) cosmology in which a canonical pseudo-Nambu–Goldstone boson (pNGB) drives late-time cosmic acceleration and controls the masses of massive neutrinos through an exponential coupling. Using current cosmic microwave background, baryon acoustic oscillation, and type Ia supernova data, we reconstruct the neutrino-mass evolution and show that the cosmologically inferred effective neutrino mass can fall below the present-day normal-ordering floor while remaining consistent with oscillation constraints. The same interaction produces a distinctive late-time signature, with 96\% of the posterior expansion histories exhibiting a crossing of the apparent phantom divide, $w_{\rm app}=-1$, at $z_{\rm cross}=0.90^{+0.28}_{-0.13}$ within 68\% C.L. The underlying scalar field nevertheless remains canonical, with $w_\phi \geq -1$. Our results therefore show that neutrino mass variation can simultaneously alleviate the apparent neutrino-mass tension and generate the phantom-crossing behavior favored by current cosmological data, without introducing a phantom field or violating the null energy condition.

\end{abstract}



\maketitle



\noindent\emph{Introduction---} Recent baryon acoustic oscillation (BAO)  measurements from the Dark Energy Spectroscopic Instrument (DESI), combined with cosmic microwave background (CMB) and type Ia supernova (SN) data, favor an evolving dark-energy equation of state (EoS) over a cosmological constant $\Lambda$ at a significance of $2.8\sigma$--$4.2\sigma$, depending on the SN sample~\cite{DESI:2025zgx}. Within $w_0w_a$ parameterizations, the preferred region, $w_0>-1$ and $w_a<0$, corresponds to an effective crossing of the phantom divide. On the other hand, the background degeneracy between such parameterizations and minimally coupled canonical scalar fields, for which $w_\phi\geq-1$, maps
a given realization of quintessence scenarios into the 
$w_0w_a$ plane, yielding the same cosmological observables~\cite{Scherrer:2015tra,Wolf:2023uno,Shlivko:2025fgv,deSouza:2025rhv}. Whether the inferred crossing can arise from a physically motivated model therefore remains an open question.

Neutrino masses present a related cosmological puzzle. Flavor-oscillation experiments require nonzero neutrino masses and imply the lower bounds $\sum m_\nu \gtrsim 0.06\;\mathrm{eV}$ and $\gtrsim 0.10\;\mathrm{eV}$ for the normal and inverted orderings, respectively~\cite{ParticleDataGroup:2024cfk}. In contrast, within the $\Lambda$CDM model, CMB+DESI DR2 BAO constrain $\sum m_\nu<0.064\;\mathrm{eV}$ at $95\%$ confidence, while recent combinations of SPT, ACT, CMB, and DESI BAO yield even tighter bounds, such as $\sum m_\nu<0.055\;\mathrm{eV}$~\cite{DESI:2025zgx,DESI:2025ejh, Camphuis_2026}. These limits fall below the normal-ordering floor and strongly disfavor the inverted ordering~\cite{Jimenez22,Gariazzo23,Jiang24}. Analyses allowing an effective neutrino-mass parameter to cross zero can even prefer vanishing or negative values~\cite{Green:2024xbb,Cozzumbo:2025ewt,Garcia-Garcia:2026nzy}. This apparent tension with oscillation results is, however, strongly model dependent. For instance, allowing $w_0w_a$ parameterizations of the dark-energy equation of state relaxes the DESI DR2 bound to $\sum m_\nu<0.129\;\mathrm{eV}$~\cite{DESI:2025ejh}, whereas physically motivated thawing quintessence models can retain constraints comparable to $\Lambda$CDM, with $\sum m_\nu < 0.067\;\mathrm{eV}$~\cite{Rodrigues:2025hso,Rodrigues:2026rgy}. This sensitivity reflects the correlations between neutrino mass, the expansion history, and structure growth~\cite{Hannestad:2005gj,RoyChoudhury:2019hls}.

Mass-varying-neutrino models provide a direct physical connection between these two problems. A coupling to the dark-energy scalar makes the neutrino mass field-dependent, $m_\nu=m_\nu(\phi)$, and allows its cosmological value to differ from its present-day value~\cite{Fardon:2003eh}. As a consequence, the apparent dark energy equation of state can cross the phantom divide without the presence of a phantom field~\cite{Antusch:2008hj}. In this {\emph{Letter}}, we investigate a MaVaN cosmology in which late-time acceleration is driven by a canonical pseudo-Nambu--Goldstone boson with a radiatively protected periodic potential~\cite{Freese:1990rb,Frieman:1995pm,Marsh:2015xka,Sengupta:2026gyg}. Through an exponential coupling normalized at the present field value, the scalar controls the massive-neutrino sector. We fix the present-day neutrino mass to $M_{\nu,0}=0.06\,\mathrm{eV}$, corresponding to the minimum summed neutrino mass implied by oscillation experiments for normal ordering~\cite{ParticleDataGroup:2024cfk}, while allowing the mass to evolve at earlier epochs. We implement the coupled background and linear perturbation equations in CLASS and constrain the model using current CMB, DESI DR2 BAO, and Type Ia supernova data. By reconstructing $M_{\nu,\mathrm{eff}}(z)$ and $w_{\rm app}(z)$, we show that the model provides a physical realization of the apparent phantom-crossing behavior suggested by current observations, while the underlying scalar field remains canonical and satisfies the null energy condition.


\vspace{0.2cm}
\noindent\emph{Neutrino--dark-energy interaction---} We assume a spatially flat FLRW background and work in reduced-Planck units $M_{\rm Pl}=1$. In this framework, the coupling between the neutrino and dark energy sectors modifies the conservation equations for each component, and 
the modified Klein-Gordon equation can be written as~\cite{Brookfield:2005td,Brookfield:2005bz, Ichiki:2007ng,daFonseca:2023ury,bundeliya2026}
\begin{equation}
\ddot{\phi} + 3H\dot{\phi} + \frac{dV}{d\phi} = -\frac{Q}{\dot{\phi}}, \label{eq:phi_eom}
\end{equation}
where \begin{equation}
    Q \equiv \frac{d\ln m_\nu(\phi)}{d\phi}\,\dot\phi\,
    (\rho_\nu-3p_\nu),
\end{equation}
the overdot denotes a derivative with respect to cosmic time, and the scalar-field energy density and pressure are defined as $\rho_\phi=\dot{\phi}^{\,2}/2+V(\phi)$ and $p_\phi=\dot{\phi}^{\,2}/2-V(\phi)$, respectively, with $V(\phi)$ corresponding to the scalar-field potential. The right-hand sides of Eq.~\ref{eq:phi_eom} are suppressed while neutrinos are relativistic ($p_\nu \simeq \rho_\nu/3$). As neutrinos become nonrelativistic ($p_\nu \ll \rho_\nu$), the coupling becomes effective. 

In our analysis, we adopt the exponential mass function,
\begin{equation}
  m_\nu(\phi)= M_{\nu,0}
  \exp\!\left[\beta(\phi-\phi_0)\right].
  \label{eq:mass-function}
\end{equation}
The coupling parameter $\beta$ determines the interaction strength. For $\beta=0$, the uncoupled quintessence limit is recovered, whereas larger values of $\beta$ enhance the energy transfer between the scalar field and neutrinos. The direction of this transfer, however, depends on the sign of $\dot{\phi}$. We further normalize the coupling relative to the present-day scalar-field value $\phi_0$, such that the parameter $M_{\nu,0}$ corresponds to the total neutrino mass today, $M_{\nu,0} \equiv \sum_i m_{\nu_i,0}$. From now on, we define 
 $ {d\ln m_\nu(\phi)}/{d\phi} = \beta$.
We adopt the pseudo-Nambu--Goldstone boson potential\footnote{The pNGB potential is very well motivated by shift-symmetric theories~\cite{Freese:1990rb,Frieman:1995pm,Marsh:2015xka}.}
\begin{equation}
V(\phi)=V_0\left[1+\cos\left(\frac{\phi}{f}\right)\right],
\label{eq:pngb-potential}
\end{equation}
where $f$ is the symmetry-breaking scale and $V_0$ sets the potential amplitude.

\vspace{0.2cm}
\noindent\emph{Intrinsic, effective, and apparent equations of state.---} Because the two sectors exchange energy, it is important to distinguish the intrinsic, effective, and apparent equations of state.  The intrinsic scalar-field EoS is
\begin{equation}
  w_\phi\equiv\frac{p_\phi}{\rho_\phi}\geq-1,
  \label{eq:wphi}
\end{equation}
where $w_\phi \geq -1$ follows from the canonical kinetic term.  One may rewrite the scalar field continuity equation as a formally uncoupled continuity equation,
\begin{equation}
  \dot\rho_\phi+3H(1+w_{\phi,{\rm eff}})\rho_\phi=0,
\end{equation}
where
\begin{equation}
  w_{\phi,{\rm eff}}
  =w_\phi+\frac{Q}{3H\rho_\phi}.
  \label{eq:weff}
\end{equation}
This effective EoS is a bookkeeping device: it describes the modified dilution rate of $\rho_\phi$.  Similarly, $w_{\nu,{\rm eff}}=w_\nu-Q/(3H\rho_\nu)$ describes the neutrino dilution rate. Neither quantity is, in general, what an observer reconstructs after assuming
constant mass, i.e., in the case of separately conserved neutrinos.

To define that reconstructed quantity, consider a reference neutrino sector with the same phase-space distribution and the same present-day mass $M_{\nu,0}$, but with this mass held constant at all redshifts. Let its energy density and pressure be $\bar\rho_\nu(a)$ and $\bar p_\nu(a)$.  An observer who subtracts this reference sector from the true total dark-sector stress tensor would assign the apparent EoS as
\begin{equation}
  w_{\rm app}(a)= \frac{p_{\rm DE}^{\rm app}}{\rho_{\rm DE}^{\rm app}}=
  \frac{p_\phi+p_\nu-\bar p_\nu}
       {\rho_\phi+\rho_\nu-\bar\rho_\nu}.
  \label{eq:wapp-general}
\end{equation}
Equation~\eqref{eq:wapp-general}, including both pressure terms, is the definition used in our numerical reconstruction.

The nonrelativistic limit provides a useful interpretation.  When
$p_\nu\simeq\bar p_\nu\simeq0$,
\begin{equation}
  \rho_\nu(a)=\rho_{\nu,0}a^{-3}
  \frac{m_\nu(\phi)}{M_{\nu,0}},
  \qquad
  \bar\rho_\nu(a)=\rho_{\nu,0}a^{-3},
\end{equation}
and Eq.~\eqref{eq:wapp-general} reduces to~\cite{Brookfield:2005bz}
\begin{align}
  w_{\rm app}&=\frac{w_\phi}{1-x},
  \label{eq:wapp-nr}\\
  x&\equiv-\frac{\rho_{\nu,0}a^{-3}}{\rho_\phi}
  \left[\frac{m_\nu(\phi)}{M_{\nu,0}}-1\right].
  \label{eq:x-definition}
\end{align}
Today $a=1$, $x=0$, and $w_{\rm app}(0)=w_\phi(0)$. If the neutrino mass was smaller in the past,
then $x>0$ and a negative $w_\phi$ is mapped to a more negative
$w_{\rm app}$. The apparent EoS may consequently cross $-1$ even though the canonical field never does.  This crossing reflects a mismatch between the true interacting model and the reference constant-mass model; it is not a ghost instability.  If $\rho_{\rm DE}^{\rm app}$ approaches zero, $w_{\rm app}$ becomes singular as a reconstructed ratio, even when the true background remains regular.


\vspace{0.2cm}
\noindent\emph{Linear perturbations.---}
Neutrino-dark energy interaction has also an important impact on the evolution of cosmological perturbations~\cite{Brookfield:2005bz, Antusch:2008hj,Afshordi_2005}.  For clarity, primes in this subsection denote conformal-time derivatives and $\mathcal H=aH$. In synchronous gauge the scalar perturbation satisfies~\cite{Brookfield:2005bz,Ichiki:2007ng,daFonseca:2023ury}
\begin{align}
  \delta\phi''+2\mathcal H\delta\phi'
  +\left(k^2+a^2V_{,\phi\phi}\right)\delta\phi
  \nonumber\\
  {}+\frac{1}{2}h'\phi'
  ={}&-a^2\beta\left(\delta\rho_\nu-3\delta p_\nu\right).
  \label{eq:delta-phi}
\end{align}
Writing the perturbed distribution as $f=f_0(q)[1+\Psi(\bm k,q,\mu,\tau)]$ and denoting its monopole by $\Psi_0$, the mass perturbation $\delta m_\nu/m_\nu=\beta\delta\phi$ contributes directly to the neutrino density and pressure perturbations:
\begin{align}
  \delta\rho_\nu={}&\frac{g_\nu}{2\pi^2a^4}
  \int_0^\infty dq\,q^2f_0(q)
  \left[
    \epsilon\Psi_0+
    \frac{a^2m_\nu^2}{\epsilon}\beta\delta\phi
  \right],
  \label{eq:delta-rho-nu}\\
  \delta p_\nu={}&\frac{g_\nu}{6\pi^2a^4}
  \int_0^\infty dq\,\frac{q^4}{\epsilon}f_0(q)
  \left[
    \Psi_0-
    \frac{a^2m_\nu^2}{\epsilon^2}\beta\delta\phi
  \right],
  \label{eq:delta-p-nu}
\end{align}
where $q=ap$ is the comoving momentum and
$\epsilon(q,\phi)=\sqrt{q^2+a^2m_\nu^2(\phi)}$ is the comoving energy.  The
coupling also modifies the dipole of the exact neutrino Boltzmann hierarchy,
schematically as
\begin{align}
  \Psi_1'={}&\frac{qk}{3\epsilon}(\Psi_0-2\Psi_2)
  -\frac{\beta k a^2m_\nu^2}{3q\epsilon}
  \frac{d\ln f_0}{d\ln q}\,\delta\phi.
  \label{eq:dipole}
\end{align}


\vspace{0.2cm}
\noindent\emph{Methodology and results.---} We implemented the complete coupled background and linear perturbation system in the Boltzmann solver CLASS~\cite{lesgourgues11,Blas:2011rf} and performed a Markov chain Monte Carlo analysis with the \textsc{Cobaya} sampler~\cite{Torrado_2021}. The baseline likelihood combines Planck 2018 low-$\ell$ temperature and polarization, Planck NPIPE CamSpec high-$\ell$ TT, TE, and EE, Planck PR4 CMB lensing, DESI DR2 BAO, and the DES five-year Dovekie-calibrated SN sample~\cite{planck18,DESI:2025zgx,desy5}. Pantheon+ provides an independent SN comparison~\cite{Brout_2022,Scolnic_2022}. Nonlinear matter corrections are modeled with \textsc{halofit}.

The neutrino sector contains one thermally populated massive species whose present-day mass is fixed to $M_{\nu,0}=0.06\,\mathrm{eV}$. We set
$N_{\rm ur}=2.0308$, which yields $N_{\rm eff}\simeq3.044$ at early times. In addition to the six standard cosmological parameters, we sample the neutrino--scalar coupling strength, $-10\leq\beta\leq10$, the pNGB symmetry-breaking scale through a logarithmic prior, $0.1\leq f_{\rm PNGB}/M_{\rm Pl}\leq3$, and the dimensionless initial misalignment angle, $0\leq\theta_{\rm ini}\equiv\phi_{\rm ini}/f_{\rm PNGB}\leq\pi$. The prior is log-uniform in $\ln f_{\rm PNGB}$, while $\beta$ and $\theta_{\rm ini}$ are assigned uniform priors. The initial field value is therefore not sampled independently, but is determined as a derived parameter, restricting the initial field displacement to the positive half-period of the pNGB potential. The symmetric interval for $\beta$ allows either direction of neutrino-mass evolution, while the broad priors on the pNGB scale and initial misalignment angle allow a wide range of scalar-field histories. The prior on $f_{\rm PNGB}$ brackets the Planck-scale value motivated in pNGB quintessence while allowing moderately super-Planckian effective scales~\cite{Frieman:1995pm}. At substantially larger $f_{\rm PNGB}$, the potential becomes increasingly flat and this scale becomes weakly identifiable; consequently, the upper endpoint specifies the phenomenological prior domain rather than a physical exclusion. The broad prior on the initial misalignment angle removes the assumption of a fixed initial field position, with $\phi_{\rm ini}=f_{\rm PNGB}\theta_{\rm ini}$, while the initial field velocity is set to zero. For each parameter point, $V_0$ is determined through a shooting procedure such that the present-day scalar density satisfies the spatial-flatness condition. The coupling is
normalized at the present scalar-field value, ensuring $M_\nu(z=0)=M_{\nu,0}$.

\begin{figure}[t]
    \centering
    \includegraphics[width=\columnwidth]{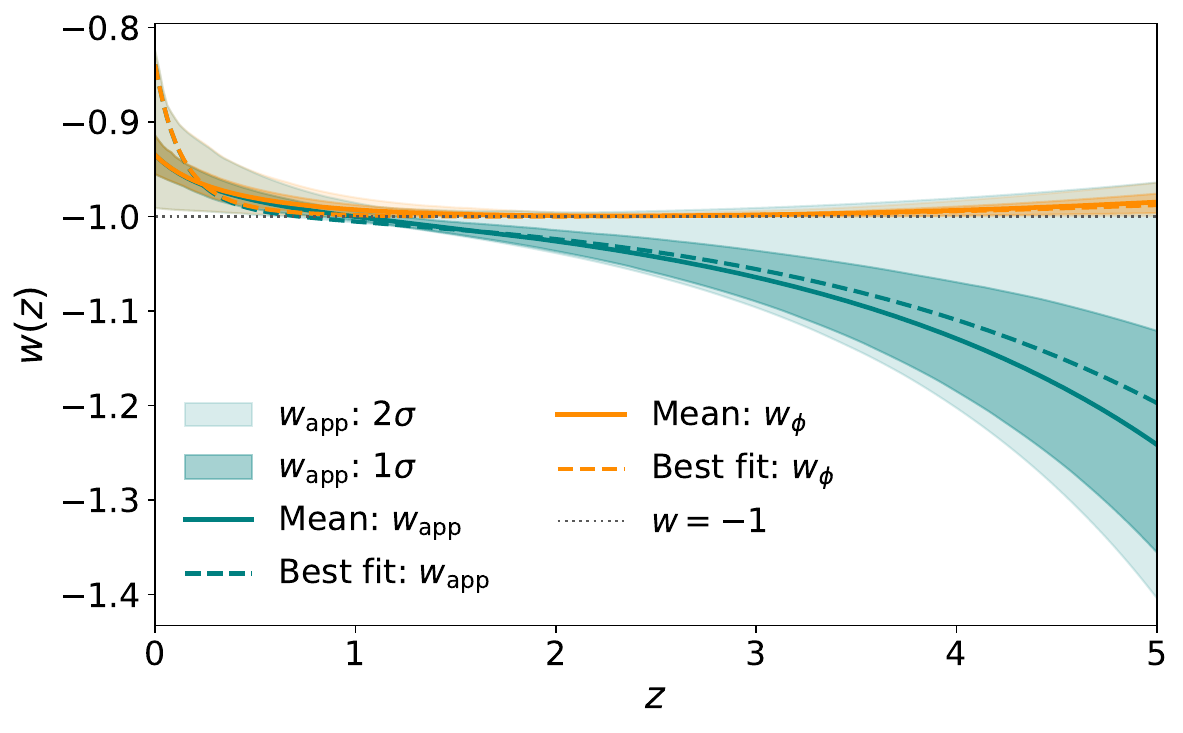}
   \caption{Reconstructed $w_{\rm app}$ (teal) and $w_\phi$ (dark orange)  for CMB $+$ DESI DR2 $+$ DES Y5-Dovekie. Solid and dashed curves denote the posterior means and best-fit histories, respectively; dark and light bands show the $68\%$ and $95\%$ credible intervals. The horizontal dotted line marks $w=-1$.}
    \label{fig:1}
\end{figure}

Figure~\ref{fig:1} displays the reconstructed apparent EoS, $w_{\rm app}$, together with the intrinsic scalar EoS, $w_\phi$, for the baseline data combination. The canonical field remains at or above $-1$ throughout, while the apparent quantity crosses the phantom divide in $96\%$ of the posterior histories. Conditional on a crossing, its median redshift and $68\%$ credible interval are $ z_{\rm cross}=0.9^{+0.28}_{-0.13}$. The best-fit history exhibits a very similar transition. The high posterior incidence shows that the crossing is not an isolated trajectory: it is a characteristic prediction of the data-favored region. Physically, a constant-mass reconstruction assigns $\rho_\nu-\bar\rho_\nu$ to dark energy, thereby generating the crossing without a phantom degree of freedom.

\begin{figure}[t]
    \centering
    \includegraphics[width=\columnwidth]{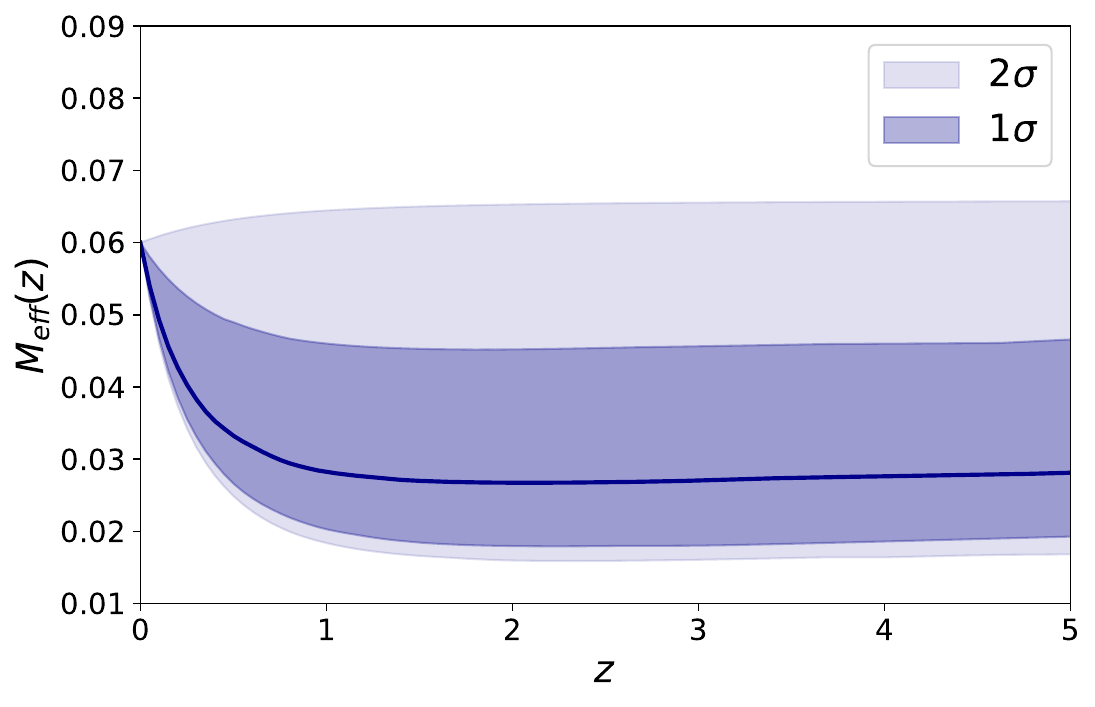}
    \caption{Reconstructed $M_{\nu,\mathrm{eff}}(z)$ for CMB $+$ DESI DR2 $+$ DES Y5-Dovekie. The central curve is the posterior median; dark and light bands show the $68\%$ and $95\%$ credible intervals.}
    \label{fig:2}
\end{figure}

Figure~\ref{fig:2} shows the effective mass history, $M_{\nu,\mathrm{eff}}(z)\equiv M_\nu[\phi(z)]$. Its posterior median lies below the present-day value at earlier epochs and evolves to $M_{\nu,\mathrm{eff}}(0)=0.06\,\mathrm{eV}$ by construction. Cosmological observables can therefore correspond to a sub-floor effective mass while the current mass remains at the normal-ordering minimum implied by terrestrial oscillation measurements. 

In the Appendix, we present the marginalized parameter constraints obtained from our analysis. The numerical results are summarized in Table~\ref{tab:constraints}, while the corresponding one- and two-dimensional marginalized posterior distributions are shown in Fig.~\ref{fig:posterior_triangle}. For the CMB + DESI DR2 + DES Y5 combination, we find $\beta = 5.4^{+3.9}_{-5.3}$, $f_{\rm PNGB}=1.1^{+1.5}_{-0.99}$, and $\phi_{ini}< 5.47$ at the $95\%$ confidence level. The data present a preference for the interacting solutions $\beta>0$; the corresponding uncoupled pNGB quintessence $\beta=0$ lies at the edge of the $95\%$ confidence level. The inferred $H_0$, $\Omega_{\rm m}$, and $\sigma_8$ remain compatible with their $\Lambda$CDM values. Replacing DES Y5 with Pantheon+ produces the same qualitative mass and EoS histories, demonstrating that the mechanism is not tied to a particular SN compilation.

We additionally perform a Bayesian model comparison using the \textsc{MCEvidence} package~\cite{Heavens:2017afc}. For the baseline dataset combination, we obtain $\Delta\ln B_{\rm MaVaN,\Lambda{\rm CDM}}=0.3$ at nearest-neighbor order $k=1$, corresponding to a Bayes factor of approximately $1.3$. Varying the nearest-neighbor order from $k=1$ to $k=4$ gives $0.3\leq\Delta\ln B\leq1.1$. Although all four estimates favor MaVaN, their magnitude indicates inconclusive evidence, with at most a weak preference for MaVaN across the tested nearest-neighbor orders, providing limited Bayesian discrimination between MaVaN and $\Lambda$CDM. This comparison is conditional on the adopted prior distributions and ranges of the additional MaVaN parameters, including the coupling strength, the pNGB scale, and the initial scalar-field configuration.

\vspace{0.2cm}
\noindent\emph{Discussion and conclusions.---} We have demonstrated that a canonical pNGB cosmology with mass-varying neutrinos reproduces two features that appear disconnected in conventional analyses. First, the reconstructed neutrino mass is smaller than $0.06\;\mathrm{eV}$ at earlier times and evolves toward $M_{\nu,\mathrm{eff}}(0)=0.06\;\mathrm{eV}$ today. Consequently, cosmological observables can respond to a mass history below the present-day normal-ordering floor without conflicting with the oscillation lower limit at the present epoch. Cosmology and terrestrial experiments consequently probe different points along the same mass history. 

Second, the canonical field satisfies $w_\phi\geq-1$ throughout its evolution, whereas $w_{\rm app}$, reconstructed under the assumption of constant-mass neutrinos, can cross the phantom divide.  The crossing occurs in $96\%$ of posterior histories, with $z_{\rm cross}=0.9^{+0.28}_{-0.13}$ among crossing histories. Its prevalence across the posterior establishes that it is not a finely tuned property of the best fit. Instead, it follows directly from analyzing a varying-mass neutrino sector as if its mass were constant. The same behavior is recovered with DES Y5-Dovekie and Pantheon+, demonstrating robustness to the SN compilation. The posterior favors interacting solutions ($\beta>0$). Therefore, across the favored region, mass-varying neutrinos provide a concrete and predictive route by which a reduced cosmological neutrino mass and apparent phantom dark energy emerge from one canonical interacting theory. The Bayesian model comparison yields $\Delta\ln B=0.3$ for the baseline dataset combination and adopted priors, corresponding to inconclusive evidence on the modified Jeffreys scale. Thus, the model comparison does not establish a clear preference for MaVaN over $\Lambda$CDM.

In ongoing work, we are extending this analysis by treating the present-day neutrino mass sum \(M_{\nu,0}\) as a free parameter, and exploring alternative scalar potentials. We will assess how these extensions affect the inferred neutrino mass history and apparent phantom crossing, and investigate whether CMB lensing and measurements of structure growth can distinguish the model from other dynamical dark energy scenarios.

\vspace{0.2cm}
\noindent\emph{Acknowledgments.---} 
GR acknowledges financial support from Conselho Nacional de Desenvolvimento Científico e Tecnológico (CNPq), through the FNDCT/MCTI, under the Programa de Capacitação Institucional (PCI/Observat\'orio Nacional). JR is supported by CNPq grant No. 406718/2025-3. SSC is financed by CNPq through the Programa de Repatriação de Talentos – Conhecimento Brasil, grant No. 446810/2024-0. JA is supported by CNPq grants (No. 307683/2022-2 and 448158/2025-6) and Funda\c{c}\~ao de Amparo \`a Pesquisa do Estado do Rio de Janeiro (FAPERJ) grant No. 299312 (2023). The development of this work was aided by the National Observatory Data Center (CPDON).

\vspace{0.2cm}
\noindent\emph{Data availability}.---
The observational data used in this work are publicly available from the sources cited in the text. The modified CLASS code, Cobaya configuration files, and analysis scripts supporting the results are available from the corresponding author upon reasonable request.

\onecolumngrid
\appendix*
\section{Marginalized parameter constraints}
\label{app:details}
\renewcommand{\theequation}{A\arabic{equation}}

\raggedbottom
\setlength{\floatsep}{8pt plus 1pt minus 1pt}
\setlength{\textfloatsep}{8pt plus 1pt minus 1pt}
\setlength{\intextsep}{8pt plus 1pt minus 1pt}

\begin{figure*}[t]
    \centering
    \includegraphics[width=0.90\textwidth]{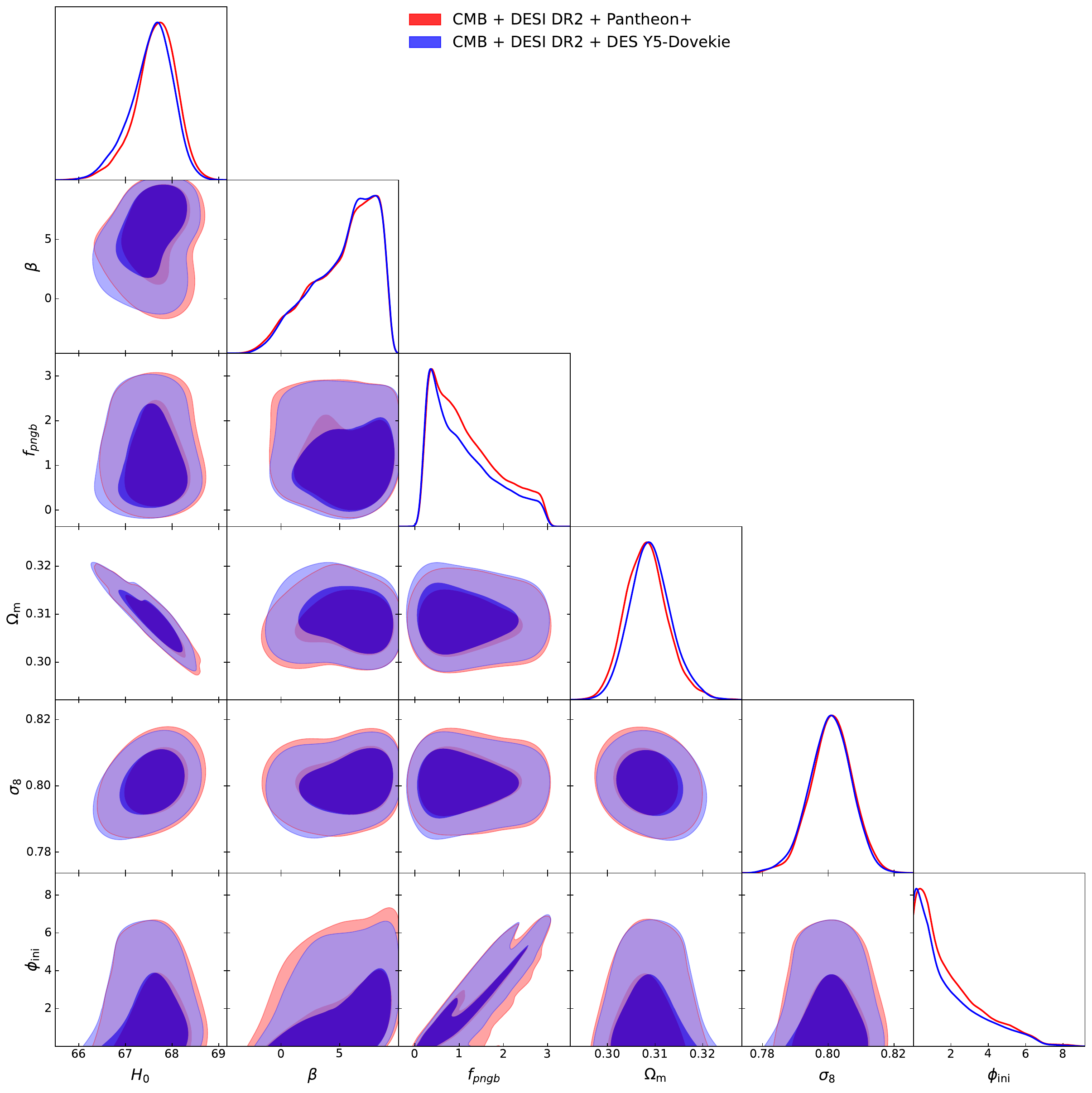}
    \caption{Marginalized posteriors for CMB $+$ DESI DR2 combined with DES Y5-Dovekie (blue) or Pantheon+ (red). Diagonal panels show the one-dimensional densities; off-diagonal panels show the $68\%$ and $95\%$ credible regions.}
    \label{fig:posterior_triangle}
\end{figure*}

Table~\ref{tab:constraints} reports selected marginalized constraints obtained from the baseline combination of CMB, DESI DR2 BAO, and DES Y5-Dovekie data, together with the corresponding Bayes factor relative to $\Lambda$CDM. The quoted intervals are obtained after marginalizing over the six standard cosmological parameters, the MaVaN parameters $\beta$, $f_{\rm PNGB}$, and $\phi_{\rm ini}$, and the complete set of foreground and calibration nuisance parameters. They therefore incorporate uncertainty in the initial scalar-field position rather than representing constraints conditional on a fixed initial configuration. Unless stated otherwise, the uncertainties correspond to 95\% credible intervals. The evolution of $\phi$ is, in turn, controlled by both $f_{\rm PNGB}$ and $\phi_{\rm ini}$, producing the correlations visible in Fig.~\ref{fig:posterior_triangle}. The posterior for $\beta$ is broad but concentrated toward positive coupling values. This behavior reflects the fact that cosmological observables do not constrain $\beta$ independently, because the coupling affects them through the resulting neutrino-mass evolution. In our model,
\begin{equation}
\ln\!\left[\frac{M_\nu(z)}{M_{\nu,0}}\right] =\beta\,[\phi(z)-\phi_0].
\label{eq:endmatter-mass}
\end{equation}

The data therefore constrain the combination of the coupling and the scalar-field excursion. A larger value of $\beta$ can be compensated by a smaller variation of $\phi$, producing similar expansion and structure-growth histories. Since the field excursion also depends on $f_{\rm PNGB}$ and $\phi_{\rm ini}$, this generates correlations among the three parameters. 

Figure~\ref{fig:posterior_triangle} compares the corresponding one- and two-dimensional marginalized posteriors obtained when DES Y5-Dovekie is replaced by Pantheon+. The substantial overlap in $H_0$, $\Omega_{\rm m}$, and $\sigma_8$ shows that the inferred standard cosmological parameters are stable against the choice of SN compilation. Although the low-redshift distance information modifies the relative posterior weight within the interacting sector, both combinations support the same qualitative physical picture: a cosmological neutrino mass that evolves toward its fixed present-day value and an apparent dark-energy equation of state capable of crossing the phantom divide without requiring a phantom scalar field.

The final entry in Table~\ref{tab:constraints} gives $\Delta\ln B\equiv\ln\mathcal{Z}_{\rm MaVaN}-\ln\mathcal{Z}_{\Lambda{\rm CDM}}$, evaluated using the same dataset combination and nuisance-parameter treatment for both models. With this convention, positive values favor the MaVaN model, whereas negative values favor $\Lambda$CDM. Because the Bayesian evidence integrates over the full parameter space, the quoted value applies specifically to the prior ranges adopted in this analysis.

\begin{table}[h]
\centering
\caption{
Marginalized constraints obtained using the baseline case. The reported uncertainties correspond to 95\% confidence intervals. 
}
\label{tab:constraints}
\begin{ruledtabular}
\begin{tabular}{lc}
Parameter & CMB +DESI DR2+DES Y5-Dovekie \\
\hline
$\beta$ & $5.4^{+3.9}_{-5.3}$  \\
$f_{\rm PNGB}$ & $1.10^{+1.5}_{-0.99} $  \\
$\phi_{ini}$ & $< 5.47 $ \\
$H_0$ & $67.55^{+0.79}_{-1.0}$  \\
$\Omega_{\rm m}$ & $0.309^{+0.0093}_{-0.0082}$ \\
$\sigma_8$ & $0.801^{+0.013}_{-0.013}$ \\
\hline
$\Delta\ln B$ & $0.3$ \
\end{tabular}
\end{ruledtabular}
\end{table}

\clearpage
\twocolumngrid
\bibliography{bibliography_cleaned}

\end{document}